\RequirePackage[OT1]{fontenc}
\documentclass{article}
\usepackage{spconf,amsmath,graphicx}
\usepackage{amsfonts}
\usepackage{cite}
\usepackage{multirow}
\usepackage[acronym]{glossaries}
\usepackage{support-caption}
\usepackage{subcaption}
\usepackage{siunitx}
\usepackage[table]{xcolor}  %
\usepackage{hyperref}
\usepackage{booktabs}

\makeatletter
\def\bstctlcite{\@ifnextchar[{\@bstctlcite}{\@bstctlcite[@auxout]}}
\def\@bstctlcite[#1]#2{\@bsphack
  \@for\@citeb:=#2\do{%
    \edef\@citeb{\expandafter\@firstofone\@citeb}%
    \if@filesw\immediate\write\csname #1\endcsname{\string\citation{\@citeb}}\fi}%
  \@esphack}
\makeatother

\DeclareSIUnit{\nothing}{\relax}

\newacronym{2d}{2D}{two-dimensional}

\newacronym{asr}{ASR}{automatic speech recognition}
\newacronym{sid}{SID}{speaker identification}
\newacronym{nmt}{NMT}{neural machine translation}
\newacronym{tts}{TTS}{text-to-speech}

\newacronym{brir}{BRIR}{binaural room impulse response}

\newacronym{dnn}{DNN}{deep neural network}
\newacronym{lstm}{LSTM}{long short-term memory}

\newacronym{ic}{IC}{interaural coherence}
\newacronym{ild}{ILD}{interaural level difference}
\newacronym{irm}{IRM}{ideal ratio mask}
\newacronym{itd}{ITD}{interaural time difference}

\newacronym{logfbe}{log-FBE}{log-filterbank energy}

\newacronym{mse}{MSE}{mean squared error}

\newacronym{pesq}{PESQ}{perceptual evaluation of speech quality}

\newacronym{relu}{ReLU}{rectified linear unit}
\newacronym{rms}{RMS}{root mean square}

\newacronym{snr}{SNR}{signal-to-noise ratio}
\newacronym{stoi}{STOI}{short-term objective intelligibility}
\newacronym{estoi}{ESTOI}{extended short-term objective intelligibility}
\newacronym{sisnr}{SISNR}{scale-invariant signal-to-noise ratio}

\newacronym{tf}{TF}{time-frequency}
\newacronym{stft}{STFT}{short-time Fourier transform}

\newacronym{zpr}{ZPR}{zero-padding rate}

\newacronym{sde}{SDE}{stochastic differential equation}
\newacronym{ve}{VE}{variance exploding}
\newacronym{vp}{VP}{variance preserving}

\newacronym{gan}{GAN}{generative adversarial network}
\newacronym{vae}{VAE}{variational autoencoder}
\newacronym{nf}{NF}{normalizing flow}

\newacronym{ema}{EMA}{exponential moving average}

\newacronym{pc}{PC}{predictor-corrector}

\newcommand*\C{\mathbb{C}}
\newcommand*\R{\mathbb{R}}
\newcommand*\E{\mathbb{E}}
\newcommand*\bx{\boldsymbol{x}}
\newcommand*\by{\boldsymbol{y}}

\newcommand*\bn{\boldsymbol{n}}
\newcommand*\bD{\boldsymbol{D}}
\newcommand*\btheta{\boldsymbol{\theta}}
\newcommand*\diff{\mathop{}\!\mathrm{d}}
\newcommand*\wiener{\boldsymbol{\omega}_t}
\newcommand*\gaussian{\mathcal{N}}
\newcommand*\cgaussian{\gaussian_{\C}}
\newcommand*\uniform[2]{\mathcal{U}(#1, #2)}
\newcommand*\score{\nabla_{\bn_t} \log p_t(\bn_t)}

\newcommand*\denoiser{\bD_{\btheta}}
\newcommand*\sigmadata{\sigma_{\mathrm{data}}}
\newcommand*\dpesq{\Delta\text{\gls{pesq}}}
\newcommand*\destoi{\Delta\text{\gls{estoi}}}
\newcommand*\dsnr{\Delta\text{\gls{snr}}}
\newcommand*\verythinspace{\mskip0.5\thinmuskip}

\title{Diffusion-based speech enhancement in matched and mismatched conditions using a Heun-based sampler}

\name{\begin{tabular}{c}Philippe Gonzalez$^1$, Zheng-Hua Tan$^2$, Jan {\O}stergaard$^2$,\\Jesper Jensen$^2$, Tommy Sonne Alstr{\o}m$^3$, Tobias May$^1$\end{tabular}}

\address{$^1$Department of Health Technology, Technical University of Denmark\\$^2$Department of Electronic Systems, Aalborg University, Denmark\\$^3$Department of Applied Mathematics and Computer Science, Technical University of Denmark}

\begin{document}
\ninept
\bstctlcite{IEEEexample:BSTcontrol}

\maketitle

\begin{abstract}
Diffusion models are a new class of generative models that have recently been applied to speech enhancement successfully.
Previous works have demonstrated their superior performance in mismatched conditions compared to state-of-the art discriminative models.
However, this was investigated with a single database for training and another one for testing, which makes the results highly dependent on the particular databases.
Moreover, recent developments from the image generation literature remain largely unexplored for speech enhancement.
These include several design aspects of diffusion models, such as the noise schedule or the reverse sampler.
In this work, we systematically assess the generalization performance of a diffusion-based speech enhancement model by using multiple speech, noise and \gls{brir} databases to simulate mismatched acoustic conditions.
We also experiment with a noise schedule and a sampler that have not been applied to speech enhancement before.
We show that the proposed system substantially benefits from using multiple databases for training, and achieves superior performance compared to state-of-the-art discriminative models in both matched and mismatched conditions.
We also show that a Heun-based sampler achieves superior performance at a smaller computational cost compared to a sampler commonly used for speech enhancement.
\end{abstract}

\begin{keywords}
Speech enhancement, diffusion models, generalization
\end{keywords}

\section{Introduction}
\label{sec:intro}

Speech enhancement aims at recovering a clean speech signal from a mixture corrupted by interfering noise and reverberation, and has applications in automatic speech recognition, speaker identification and noise reduction for communication devices such as hearing aids.
Learning-based algorithms have dominated recent developments in speech enhancement due to their superior performance over traditional statistical methods~\cite{wang2018supervised}.
These algorithms can be categorized into discriminative and generative approaches.
Discriminative approaches learn a direct mapping from the noisy speech signal to the clean speech signal by minimizing a point-wise distance during training.
On the other hand, generative approaches learn a probability distribution over clean speech and can generate different samples from the same noisy speech input.
While it is well known that the performance of discriminative models substantially decreases in acoustic conditions that were not seen during training~\cite{gonzalez2023assessing}, the generalization potential of generative models is largely unexplored.

Score-based generative models or diffusion models~\cite{sohl2015deep,ho2020denoising,song2021score} have shown outstanding performance in the fields of image generation~\cite{ho2020denoising,song2021score,dhariwal2021diffusion,rombach2022high}, audio generation~\cite{kong2021diffwave,popov2021grad,liu2023audioldm} and video generation~\cite{ho2022video}, and have recently been applied to speech enhancement successfully~\cite{lu2021study,lu2022conditional,welker2022speech,richter2023speech,yen2023cold,wang2023cross,chen2023metric}.
Several studies have evaluated the generalization potential of diffusion-based approaches to unseen acoustic conditions.
In~\cite{lu2022conditional}, a diffusion-based speech enhancement system was trained on Voicebank+DEMAND~\cite{valentini2016speech} and tested on CHiME-4~\cite{vincent2017analysis} to simulate mismatched conditions.
In~\cite{richter2023speech}, the model was trained on Voicebank+DEMAND and tested on a dataset created from the Wall Street Journal (WSJ0) dataset~\cite{paul1992design} and CHiME-3~\cite{barker2015third}.
In~\cite{chen2023metric}, the system was trained on Voicebank+DEMAND and tested on TIMIT~\cite{garofolo1993timit} and two unseen noise types.
In all those studies, the diffusion model achieved superior performance in mismatched conditions compared to state-of-the-art discriminative systems, thus suggesting that generative approaches generalize better to unseen acoustic conditions.

However, the studies in~\cite{lu2022conditional,richter2023speech,chen2023metric} have only used one database for training and another one for testing.
As a consequence, the results are highly dependent on the particular choice of databases and are likely to change if different databases were used.
It was also shown that the generalization of state-of-the-art discriminative systems substantially improves when they are trained on multiple databases~\cite{gonzalez2023assessing}, but this has not yet been investigated for generative approaches.
Moreover, recent studies in image generation literature have proposed alternative noise schedules for the diffusion model such as cosine noise schedules~\cite{nichol2021improved,hoogeboom2023simple,kingma2023understanding}, as well as different samplers such as a 2\textsuperscript{nd} order Heun-based sampler~\cite{karras2022elucidating}, both of which remain largely unexplored for speech enhancement.

In the present work, we build upon our previous study on generalization assessment~\cite{gonzalez2023assessing} to more accurately estimate the generalization performance of a diffusion-based speech enhancement system to unseen acoustic conditions.
Specifically, we use an ensemble of five speech corpora, five noise databases and five \gls{brir} databases to train and test the system in a cross-validation manner.
For each fold, the model is trained with one or multiple databases and tested with the held out ones to estimate the performance in mismatched conditions.
By averaging the results across folds, the dependency of the results on specific databases is reduced.
We also experiment with a shifted-cosine noise schedule and two different samplers, namely the \gls{pc} sampler from~\cite{song2021score}, which has been extensively used for speech enhancement~\cite{welker2022speech,richter2023speech,lemercier2023analysing}, and the Heun-based sampler from~\cite{karras2022elucidating}, which has not yet been used in the context of speech enhancement.
To use the Heun-based sampler, we adopt a diffusion model formulation that differs from previous adaptations to speech enhancement.

\section{Proposed diffusion-based speech enhancement system}
\label{sec:diffusion}

We consider a signal model ${\by = \bx_0 + \bn_0 \in \C^{d}}$ where ${\by}$, ${\bx_0}$ and ${\bn_0}$ are complex \gls{stft} representations of the noisy speech signal, the clean speech signal and the environmental noise respectively, flattened into ${d=KT}$-dimensional vectors where ${K}$ is the number of frequency bins and ${T}$ is the number of frames.
The diffusion model progressively transforms clean speech data ${\bx_0}$ into Gaussian noise and can be trained to undo this process while conditioned on some noisy speech ${\by}$ to generate new samples from random noise realizations.
To model the diffusion process, previous studies~\cite{welker2022speech,richter2023speech} have adopted a multivariate \gls{sde} of the following general form,
\begin{equation}
  \diff \bx_t = f(t) (\bx_t - \by) \diff t + g(t) \diff \wiener,
  \label{eq:sde_xt}
\end{equation}
where ${t \in [0, T]}$ is a continuous variable indexing the diffusion process, ${f(t)(\bx_t - \by) \in \R^d}$ is the \textit{drift} coefficient, ${g(t) \in \R}$ is the \textit{diffusion} coefficient and ${\wiener \in \C^d}$ is a complex-valued standard Wiener process.
Note that ${\bx_t}$ is conditioned on ${\by}$ and this conditioning is omitted to alleviate notation.
The drift coefficient in Eq.~\eqref{eq:sde_xt} cannot be written in the form ${f(t)\verythinspace\bx_t}$ as in~\cite{karras2022elucidating}, so we use a change of variable and consider instead the \gls{sde} satisfied by the environmental noise ${\bn_t = \bx_t - \by}$,
\begin{equation}
  \diff \bn_t = f(t) \verythinspace \bn_t \diff t + g(t) \diff \wiener,
  \label{eq:sde_nt}
\end{equation}
where ${\bn_t}$ is also conditioned on ${\by}$.
This allows us to formulate the problem according to the framework laid in~\cite{karras2022elucidating}.
Specifically, it can be shown that the diffusion process ${\bn_t}$ follows the following complex-valued Gaussian conditional probability density function,
\begin{equation}
  p_{0t}(\bn_t|\bn_0) = \cgaussian \big( \bn_t ; s(t) \verythinspace \bn_0, s(t)^2 \sigma(t)^2 \verythinspace \mathbf{I} \verythinspace \big),
\end{equation}
where
\begin{equation}
  s(t) = \exp \int_0^t f(\xi) \diff \xi
  \quad \text{and} \quad
  \sigma(t)^2 = \int_0^t \frac{g(\xi)^2}{s(\xi)^2} \diff \xi.
\end{equation}
The associated reverse-time \gls{sde} is~\cite{anderson1982reverse}
\begin{equation}
  \diff \bn_t = \left[ f(t) \verythinspace \bn_t - g(t)^2 \verythinspace \score \right] \diff t + g(t) \diff \wiener,
  \label{eq:reverse_sde}
\end{equation}
where ${\score}$ is the \textit{score} function, which cannot be expressed in closed form as it involves marginalizing and integrating over the distribution of the training data ${\bn_0}$.
Nonetheless, a score model can be fitted as in~\cite{karras2022elucidating} by training a denoiser model ${\denoiser}$ to minimize a weighted ${\mathcal{L}_2}$-loss between the training data ${\bn_0}$ and the output from the diffused data ${\bn_t}$,
\begin{equation}
  \E_{\bn_0, \by} \E_{t, \bn_t | \bn_0} \Big[ w(t) \left\| \denoiser(\bn_t / s(t), \by, t) - \bn_0 \right\|_2^2 \Big],
\end{equation}
where ${t \sim \uniform{0}{T}}$ and ${w(t) \in \R}$ is a weighting function.
The expression of ${\denoiser}$ as a function of the raw neural network layers is termed \textit{preconditioning}~\cite{karras2022elucidating}.
The score is then approximated as
\begin{equation}
  \score \simeq \frac{\denoiser(\bn_t / s(t), \by, t) - \bn_t / s(t)}{s(t)\verythinspace\sigma(t)^2}.
\end{equation}
The reverse-time diffusion in Eq.~\eqref{eq:reverse_sde} can thereafter be numerically integrated starting from a random ${\bn_T \sim \cgaussian(\mathbf{0}, s(T)^2\sigma(T)^2\mathbf{I})}$ to generate an estimate of the environmental noise ${\bn_0}$, which can finally be subtracted from the noisy speech ${\by}$ to perform speech enhancement.
Note that adopting the \gls{sde} in Eq.~\eqref{eq:sde_nt} satisfied by ${\bn_t}$ to obtain a drift coefficient of the form ${f(t)\verythinspace\bx_t}$ allows us to use the Heun-based sampler from~\cite{karras2022elucidating}, which is not possible with the \gls{sde} in Eq.~\eqref{eq:sde_xt} satisfied by ${\bx_t}$ and considered in~\cite{welker2022speech,richter2023speech}.

The joint parametrization of ${f(t)}$, ${g(t)}$, ${s(t)}$ and ${\sigma(t)}$ is commonly referred to as the \textit{noise schedule}.
We choose a shifted-cosine noise schedule, as this has been shown to provide superior performance in image generation compared to linear noise schedules~\cite{nichol2021improved,hoogeboom2023simple,kingma2023understanding}.
This noise schedule is defined under the variance-preserving assumption~\cite{song2021score} in terms of the log-\gls{snr} of the diffusion process ${\lambda(t) = - 2 \log \sigma(t)}$ as
\begin{equation}
  \lambda(t) = -2 \log \left( \tan \frac{\pi t}{2} \right) + 2 \nu,
\end{equation}
where ${\nu}$ controls the center of the log-\gls{snr} distribution.
The corresponding ${f(t)}$, ${g(t)}$, ${s(t)}$ and ${\sigma(t)}$ can be derived using Eqs.~(98\nobreakdash--101) in~\cite{kingma2023understanding} to obtain ${f(t) = - \frac{1}{2} \beta(t)}$, ${g(t) = \sqrt{\beta(t)}}$, ${s(t) = \frac{1}{\sqrt{1 + \sigma(t)^2}}}$ and ${\sigma(t) = e^{-\nu}\tan \frac{\pi t}{2}}$, where
\begin{equation}
  \beta(t) = \frac{\pi}{\cos^2\!\frac{\pi t}{2}} \cdot \frac{\tan\!\frac{\pi t}{2}}{e^{2\nu} + \tan^2\!\frac{\pi t}{2}} = \frac{2 \pi \csc (\pi t)}{1 + e^{2\nu}\tan^{-2}\!\frac{\pi t}{2}}.
\end{equation}

\section{Generalization assessment}
\label{sec:generalization}

We build upon our previous work~\cite{gonzalez2023assessing} to assess the generalization of the diffusion-based speech enhancement system.
Namely, we consider an ensemble of five speech corpora, five noise databases and five \gls{brir} databases to train and test the model in a cross-validation manner.
For the speech, we use TIMIT~\cite{garofolo1993timit}, LibriSpeech (100-hour version)~\cite{panayotov2015librispeech}, WSJ SI-84~\cite{paul1992design}, Clarity~\cite{cox2022clarity} and VCTK~\cite{veaux2013voice}.
For the noise, we use TAU~\cite{heittola2019tau}, NOISEX\cite{varga1993noisex}, ICRA~\cite{dreschler2001icra}, DEMAND~\cite{thiemann2013demand} and ARTE~\cite{weisser2019ambisonic}.
Finally for the \glspl{brir}, we use Surrey~\cite{hummersone2010surrey}, ASH~\cite{shanon2021ash}, BRAS~\cite{brinkmann2021bras}, CATT~\cite{catt_brirs} and AVIL~\cite{mccormack2020higher}.
For each fold, ${N=1}$ or ${N=4}$ speech corpora, noise databases and \gls{brir} databases are used for training, while the remaining databases are held out to simulate a mismatched acoustic condition.
The ${N=1}$ case represents a \textit{low-diversity} training condition, while the ${N=4}$ case represents a \textit{high-diversity} training condition.
The results are averaged across five folds in both matched and mismatched conditions to accurately assess the generalization performance and investigate the benefit of the diversity of the training data.

\section{Experimental setup}
\label{sec:experimental}

\subsection{Mixture generation and pre-processing}
\label{subsec:mixture}

Each mixture is sampled at \SI{16}{\kilo\hertz} and generated by convolving one clean speech utterance and up to three noise segments with \glspl{brir} measured from different and random spatial locations between \SI{-90}{\degree} and \SI{90}{\degree} in the same room.
The target signal is defined as the direct-sound part of the speech signal including early reflections up to a boundary of \SI{50}{\milli\second}, as this has been shown to be beneficial for speech intelligibility~\cite{roman2013speech}.
The environmental noise thus includes late speech reflections.
Both signals are mixed at a random \gls{snr} uniformly distributed between \SI{-5}{\decibel} and \SI{10}{\decibel} and averaged across left and right channels.
For the speech, \SI{80}{\percent} of the utterances in each corpus is reserved for training and \SI{20}{\percent} for testing.
For the noise, \SI{80}{\percent} of each recording is reserved for training and \SI{20}{\percent} for testing.
For the \glspl{brir}, every other \gls{brir} of each room is reserved for training and the rest for testing.
The training and test dataset sizes in each fold are set to \SI{10}{\hour} and \SI{1}{\hour} respectively.
Note that the training dataset size is not scaled with the number of training databases ${N}$ in order to investigate the effect of the diversity of the training data.

We use a \gls{stft} frame length of 512 samples, which corresponds to \SI{32}{\milli\second} at \SI{16}{\kilo\hertz}, with a step size of 128 samples and a Hann window.
The Nyquist component is discarded to obtain ${K=256}$ frequency bins.
We apply the same transformation to the \gls{stft} coefficients as in~\cite{welker2022speech,richter2023speech} to account for their heavy-tailed distribution,
\begin{equation}
  \tilde c = A |c|^\alpha e^{i \angle c},
  \label{eq:stft_transform}
\end{equation}
where ${c}$ is the raw \gls{stft} coefficient, ${\tilde c}$ is the transformed \gls{stft} coefficient, ${A=0.15}$ and ${\alpha=0.5}$.

\subsection{SDE, preconditioning and sampler}
\label{subsec:sde}

To parametrize the \gls{sde}, we set ${T=1}$ and clamp ${\lambda(t) \geq \lambda_{\min}}$ and ${\beta(t) \leq \beta_{\max}}$ to prevent instabilities around ${t=1}$.
We choose ${\nu = 1.5}$, ${\lambda_{\min} = -12}$ and ${\beta_{\max} = 10}$ empirically after preliminary experiments.
We use the same preconditioning and loss weighting ${w(t)}$ as in~\cite{karras2022elucidating} with ${\sigmadata = 0.1}$, as this corresponds to the variance of complex \gls{stft} coefficients in our training data after applying the transformation Eq.~\eqref{eq:stft_transform}.
For integrating the reverse process, we either use the \gls{pc} sampler from~\cite{song2021score} as in~\cite{welker2022speech,richter2023speech}, or the 2\textsuperscript{nd} order Heun-based stochastic sampler from~\cite{karras2022elucidating}, which we denote as EDM.
In both cases, the time axis is discretized into ${n_{\mathrm{steps}}}$ evenly spaced steps between ${t=T}$ and ${t=0}$.
The ${n_{\mathrm{steps}}}$ parameter, which controls the number neural network evaluations and thus inference time, is varied in powers of 2 between 4 and 64.
For the \gls{pc} sampler, we use ${r=0.5}$ with a single corrector step as suggested in~\cite{richter2023speech}.
This ensures the number of neural network evaluations is the same with both samplers for a fixed ${n_{\mathrm{steps}}}$.
For the EDM sampler, we use ${S_{\mathrm{noise}}=1}$, ${S_{\min}=0}$, ${S_{\max}=\infty}$ and ${S_{\mathrm{churn}}=\infty}$, as this gave the best performance during preliminary experiments.

\subsection{Neural network and training}
\label{subsec:neural}

We use the NCSN++M neural network architecture from~\cite{lemercier2023analysing}.
This is a \SI{27.8}{\mega\nothing}-parameter version of the NCSN++ neural network initially proposed in~\cite{song2021score}.
We train the system on full-length mixtures for 100 epochs with the Adam optimizer~\cite{kingma2015adam} and a learning rate of ${1e^{-4}}$.
A bucket batching strategy is adopted to create batches of variable-lenth mixtures~\cite{gonzalez2023batching}, using 10 buckets and a dynamic batch size of \SI{32}{\second}.
Exponential moving average of the neural network weights is applied with a decay rate of 0.999.
We sample ${t \sim \uniform{t_\epsilon}{1}}$ during training with ${t_\epsilon=0.01}$ to avoid instabilities for very small ${t}$ values as previously done in literature~\cite{song2021score,richter2023speech}.

\subsection{Objective metrics and baselines}
\label{subsec:objective}

We measure the speech enhancement performance in terms of \gls{pesq}~\cite{recommendation2001perceptual}, \gls{estoi}~\cite{jensen2016algorithm} and \gls{snr}.
The results are reported in terms of average objective metric improvement between the unprocessed input mixture and the enhanced output signal.
The improvements are denoted as ${\dpesq}$, ${\destoi}$ and ${\dsnr}$ respectively.
We compare the results with discriminative systems previously investigated in~\cite{gonzalez2023assessing}, namely Conv-TasNet~\cite{luo2019conv}, DCCRN~\cite{hu2020dccrn} and MANNER~\cite{park2022manner}.
These models have \SI{4.9}{\mega\nothing}, \SI{3.7}{\mega\nothing} and \SI{21.2}{\mega\nothing} parameters respectively.
We also compare the results with SGMSE+M~\cite{lemercier2023analysing}, which is a diffusion-based speech enhancement model with the same NCSN++M neural network architecture and the same \gls{pc} sampler as our proposed model, but which uses a different noise schedule, preconditioning and loss weighting.
We train SGMSE+M as described in Sect.~\ref{subsec:neural}.

\section{Results}
\label{sec:results}

\begin{figure}
  \captionsetup{skip=5pt}
  \centering
  \includegraphics{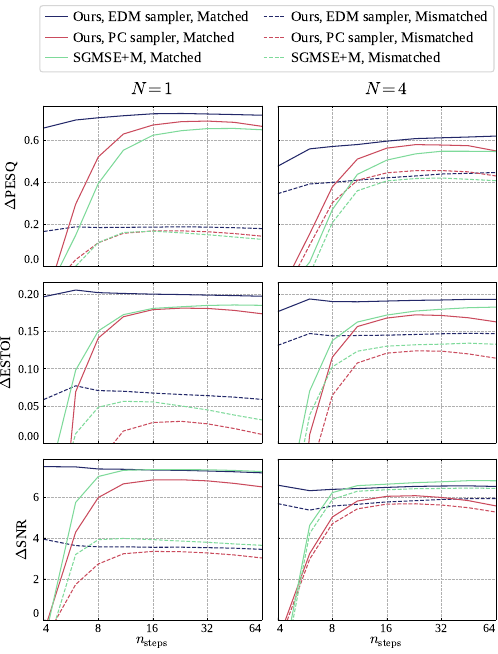}
  \caption{Average ${\dpesq}$, ${\destoi}$ and ${\dsnr}$ scores as a function of the number of sampling steps ${n_{\mathrm{steps}}}$ when training with ${N=1}$ or ${N=4}$ speech corpora, noise databases and \gls{brir} databases.}
  \label{fig:results}
\end{figure}

\begin{table*}[t]
\centering
\scriptsize
\captionsetup[subtable]{belowskip=6pt}
\captionsetup{skip=0pt}
\sisetup{table-format=1.2,detect-all}
\begin{subtable}{.49\linewidth}
\centering
\begin{tabular}{cc*{3}{S}}
\toprule
 &  & {$\Delta$PESQ} & {$\Delta$ESTOI} & {$\Delta$SNR} \\
\midrule
\multirow{9}{*}{\rotatebox[origin=c]{90}{Matched}} & Conv-TasNet & 0.63 & 0.19 & \bfseries 8.58 \\
 & DCCRN & 0.41 & 0.12 & 7.11 \\
 & MANNER & 0.68 & 0.17 & 7.24 \\
 & SGMSE+M, $n_\mathrm{steps}\!\!=\!\!16$ & 0.62 & 0.18 & 7.34 \\
 & SGMSE+M, $n_\mathrm{steps}\!\!=\!\!32$ & 0.65 & 0.18 & 7.33 \\
\cmidrule{2-5}
 & Ours, PC sampler, $n_\mathrm{steps}\!\!=\!\!16$ & 0.67 & 0.18 & 6.84 \\
 & Ours, PC sampler, $n_\mathrm{steps}\!\!=\!\!32$ & 0.69 & 0.18 & 6.79 \\
 & Ours, EDM sampler, $n_\mathrm{steps}\!\!=\!\!16$ & \bfseries 0.72 & \bfseries 0.20 & 7.32 \\
 & Ours, EDM sampler, $n_\mathrm{steps}\!\!=\!\!32$ & 0.72 & 0.20 & 7.27 \\
\midrule
\multirow{9}{*}{\rotatebox[origin=c]{90}{Mismatched}} & Conv-TasNet & 0.12 & 0.01 & 3.24 \\
 & DCCRN & 0.11 & 0.02 & 3.25 \\
 & MANNER & 0.17 & 0.04 & 3.40 \\
 & SGMSE+M, $n_\mathrm{steps}\!\!=\!\!16$ & 0.17 & 0.06 & \bfseries 3.92 \\
 & SGMSE+M, $n_\mathrm{steps}\!\!=\!\!32$ & 0.15 & 0.04 & 3.79 \\
\cmidrule{2-5}
 & Ours, PC sampler, $n_\mathrm{steps}\!\!=\!\!16$ & 0.17 & 0.03 & 3.34 \\
 & Ours, PC sampler, $n_\mathrm{steps}\!\!=\!\!32$ & 0.17 & 0.03 & 3.27 \\
 & Ours, EDM sampler, $n_\mathrm{steps}\!\!=\!\!16$ & 0.19 & \bfseries 0.07 & 3.54 \\
 & Ours, EDM sampler, $n_\mathrm{steps}\!\!=\!\!32$ & \bfseries 0.19 & 0.06 & 3.52 \\
\bottomrule
\end{tabular}
\caption{$N=1$}
\label{tab:results_a}
\end{subtable}
\begin{subtable}{.49\linewidth}
\centering
\begin{tabular}{cc*{3}{S}}
\toprule
 &  & {$\Delta$PESQ} & {$\Delta$ESTOI} & {$\Delta$SNR} \\
\midrule
\multirow{9}{*}{\rotatebox[origin=c]{90}{Matched}} & Conv-TasNet & 0.44 & 0.16 & \bfseries 7.38 \\
 & DCCRN & 0.34 & 0.12 & 6.52 \\
 & MANNER & 0.52 & 0.15 & 6.23 \\
 & SGMSE+M, $n_\mathrm{steps}\!\!=\!\!16$ & 0.50 & 0.17 & 6.63 \\
 & SGMSE+M, $n_\mathrm{steps}\!\!=\!\!32$ & 0.55 & 0.18 & 6.74 \\
\cmidrule{2-5}
 & Ours, PC sampler, $n_\mathrm{steps}\!\!=\!\!16$ & 0.56 & 0.17 & 6.03 \\
 & Ours, PC sampler, $n_\mathrm{steps}\!\!=\!\!32$ & 0.58 & 0.17 & 5.98 \\
 & Ours, EDM sampler, $n_\mathrm{steps}\!\!=\!\!16$ & 0.59 & 0.19 & 6.47 \\
 & Ours, EDM sampler, $n_\mathrm{steps}\!\!=\!\!32$ & \bfseries 0.61 & \bfseries 0.19 & 6.54 \\
\midrule
\multirow{9}{*}{\rotatebox[origin=c]{90}{Mismatched}} & Conv-TasNet & 0.29 & 0.09 & 5.77 \\
 & DCCRN & 0.23 & 0.07 & 5.42 \\
 & MANNER & 0.38 & 0.10 & 5.61 \\
 & SGMSE+M, $n_\mathrm{steps}\!\!=\!\!16$ & 0.41 & 0.13 & 6.35 \\
 & SGMSE+M, $n_\mathrm{steps}\!\!=\!\!32$ & 0.42 & 0.13 & \bfseries 6.42 \\
\cmidrule{2-5}
 & Ours, PC sampler, $n_\mathrm{steps}\!\!=\!\!16$ & 0.44 & 0.12 & 5.65 \\
 & Ours, PC sampler, $n_\mathrm{steps}\!\!=\!\!32$ & \bfseries 0.46 & 0.12 & 5.61 \\
 & Ours, EDM sampler, $n_\mathrm{steps}\!\!=\!\!16$ & 0.42 & 0.15 & 5.76 \\
 & Ours, EDM sampler, $n_\mathrm{steps}\!\!=\!\!32$ & 0.44 & \bfseries 0.15 & 5.88 \\
\bottomrule
\end{tabular}
\caption{$N=4$}
\label{tab:results_b}
\end{subtable}
\caption{Average $\dpesq$, $\destoi$ and $\dsnr$ scores in matched and mismatched conditions when training with (a) $N=1$ or (b) $N=4$ speech corpora, noise databases and \gls{brir} databases.}
\label{tab:results}
\end{table*}

Figure~\ref{fig:results} shows the average ${\dpesq}$, ${\destoi}$ and ${\dsnr}$ for the proposed diffusion-based speech enhancement system and SGMSE+M as a function of the number of sampling steps ${n_{\mathrm{steps}}}$, when training with low-diversity (${N=1}$) or high-diversity (${N=4}$) datasets and in both matched (solid lines) and mismatched (dashed lines) conditions.
The performance of the proposed model is consistently higher when using the EDM sampler compared to the \gls{pc} sampler across all metrics and configurations.
The performance benefit is largest when few sampling steps ${n_{\mathrm{steps}}}$ are used.
The EDM sampler shows good performance for ${{n_{\mathrm{steps}}=4}}$ and does not seem to benefit from increasing ${n_{\mathrm{steps}}}$ further.
On the other hand, the \gls{pc} sampler does not show an improvement in terms of any of the metrics at ${n_{\mathrm{steps}}=4}$, and substantially benefits from increasing ${n_{\mathrm{steps}}}$.
This means that the EDM sampler is able to achieve better performance at a smaller computational cost compared to the \gls{pc} sampler.
Comparing with SGMSE+M, the same trend can be observed, since SGMSE+M also uses the \gls{pc} sampler.
The proposed system with the EDM sampler outperforms SGMSE+M in terms of ${\dpesq}$ and ${\destoi}$ in all configurations, and is only outperformed in terms of ${\dsnr}$ at large ${n_{\mathrm{steps}}}$.
Finally, when using the \gls{pc} sampler, the proposed system shows superior performance compared to SGMSE+M only in terms of ${\dpesq}$.
This indicates that the considered shifted-cosine noise schedule, preconditioning and loss weighting perform similarly to their counterparts in~\cite{lemercier2023analysing}.

Table~\ref{tab:results} shows the results from the proposed diffusion-based speech enhancement system for ${n_{\mathrm{steps}}\in\{16,32\}}$ together with the discriminative baselines and SGMSE+M.
The proposed system outperforms the baseline models in terms of ${\dpesq}$ and ${\destoi}$ when training with both low-diversity (${N=1}$) and high-diversity (${N=4}$) datasets and in both matched and mismatched conditions, while Conv-TasNet and SGMSE+M show the best performance in terms of ${\dsnr}$ in matched and mismatched conditions respectively.
The substantial performance advantage achieved by Conv-TasNet in matched conditions is related to it being directly trained to minimize a \gls{snr}-based loss.
All models show a performance drop in matched conditions when going from ${N=1}$ to ${N=4}$, which can be explained by the fact that they optimize a wider range of acoustic conditions with the same number of learnable parameters.
However, the performance in mismatched conditions substantially increases for all models.
This highlights the benefit of using multiple databases during training for generalization and is in line with~\cite{gonzalez2023assessing}.

\section{Conclusion}
\label{sec:conclusion}

In the present work, we investigated the performance of a diffusion-based speech enhancement system in matched and mismatched conditions by using multiple speech, noise and \gls{brir} databases in a cross-validation manner.
The proposed diffusion-based model builds upon recent developments from image generation literature and uses a shifted-cosine noise schedule and the EDM sampler, which both have not been used in the context of speech enhancement before.
In order to use the EDM sampler, we considered the \gls{sde} satisfied by the environmental noise signal instead of the clean speech signal, such that the diffusion model can be formulated according to~\cite{karras2022elucidating}.
We showed that the proposed system substantially benefits from using multiple databases during training, and achieves superior performance compared to state-of-the-art discriminative systems in both matched and mismatched conditions.
We also showed that the EDM sampler achieves superior performance at a smaller number of sampling steps compared to the \gls{pc} sampler, thus reducing the computational cost.
Future work will investigate the effect of the training dataset size on generalization, as well as other design aspects of the diffusion model such as the preconditioning or the amount of stochasticity during sampling.

\bibliographystyle{IEEEbib}
\bibliography{IEEEabrv,abbrv,refs}

\end{document}